\documentclass{PoS}

\title{Higgs production within $k_t$-factorization \\
with unintegrated gluon distributions}

\ShortTitle{Higgs production within $k_t$-factorization}

\author {\speaker{Antoni Szczurek\thanks{Also at University of Rzesz\'ow, 35-959 Rzesz\'ow, Poland.}}\\
        The H. Niewodnicza\'{n}ski Institute of Nuclear Physics Polish Academy of Sciences,\\
        ul. Radzikowskiego 152, 31-342 Krak\'ow, Poland\\
        E-mail: \email{Antoni.Szczurek@ifj.edu.pl}}

\abstract{
We present differential cross sections for Higgs boson and/or 
two-photon production from intermediate (virtual) Higgs boson within
the formalism of $k_t$-factorization. 
Resulting distributions for two photons from the Higgs boson 
are compared with recent ATLAS collaboration data.
In contrast to a recent calculation the leading order $g g \to H$ 
contribution is rather small compared to the ATLAS experimental data
($\gamma \gamma$ transverse momentum and rapidity distributions).
We include also higher-order contribution 
$g g \to H (\to \gamma \gamma) g$, $g g \to g H g$ and the contribution 
of the $W^+ W^-$ and $Z^0 Z^0$ exchanges. 
The $gg\to Hg$ mechanism gives a similar contribution as the $gg\to H$ 
mechanism.
We argue that there is almost no double counting when adding 
$gg\to H$ and $gg\to Hg$ contributions due to different topology 
of corresponding Feynman diagrams.
The final sum is comparable with the ATLAS two-photon data. 
}

\FullConference{The European Physical Society Conference on High Energy Physics\\
		22--29 July 2015\\
		Vienna, Austria}

\begin{document}

\section{Introduction}

The Higgs boson was discovered at the LHC \cite{Higgs_discovery}. 
It has been observed in a few decay channels.
The $\gamma \gamma$ and $Z^0 Z^{0,*}$ are particularly spectacular
\cite{Aad:2013wqa,Chatrchyan:2013mxa,Khachatryan:2014iha,Aad:2014aba}.


After the discovery understanding the rapidity and transverse 
momentum distributions is particularly interesting. 
While the total cross section is well under control and was calculated 
in leading-order (LO), next-to-leading order (NLO) and 
even next-to-next-to-leading order (NNLO) approximation \cite{NNLO} 
the distribution in the Higgs boson transverse
momentum is more chalanging. 

 
It was advocated recently that precise differential data for Higgs boson
in the two-photon final channel could be very useful to test and explore
unintegrated gluon distribution functions (UGDFs) \cite{Jung2013}. 
It was claimed very recently \cite{LMZ2014} that the $k_t$-factorization 
formalism with commonly used UGDFs (Kimber-Martin-Ryskin (KMR)
\cite{KMR} and Jung CCFM \cite{Jung}) gives a reasonable description of 
recent ATLAS data obtained at $\sqrt{s} =$ 8 TeV \cite{ATLAS_Higgs}. 
We perform similar calculation and draw rather different conclusions.

Here we report on our results from Ref.\cite{SLM2014} obtained within
$k_t$-factorization approach where we presented several 
differential distributions for the Higgs boson and photons from 
the Higgs boson decay at $\sqrt{s}$ = 8 TeV for various 
UGDFs from the literature.

There we have included both leading-order and next-to-leading order 
contributions.
We shall critically discuss uncertainties and open problems
in view of the recent ATLAS data.

\section{A sketch of the formalism}

\begin{figure}
\begin{center}
\includegraphics[width=5.5cm]{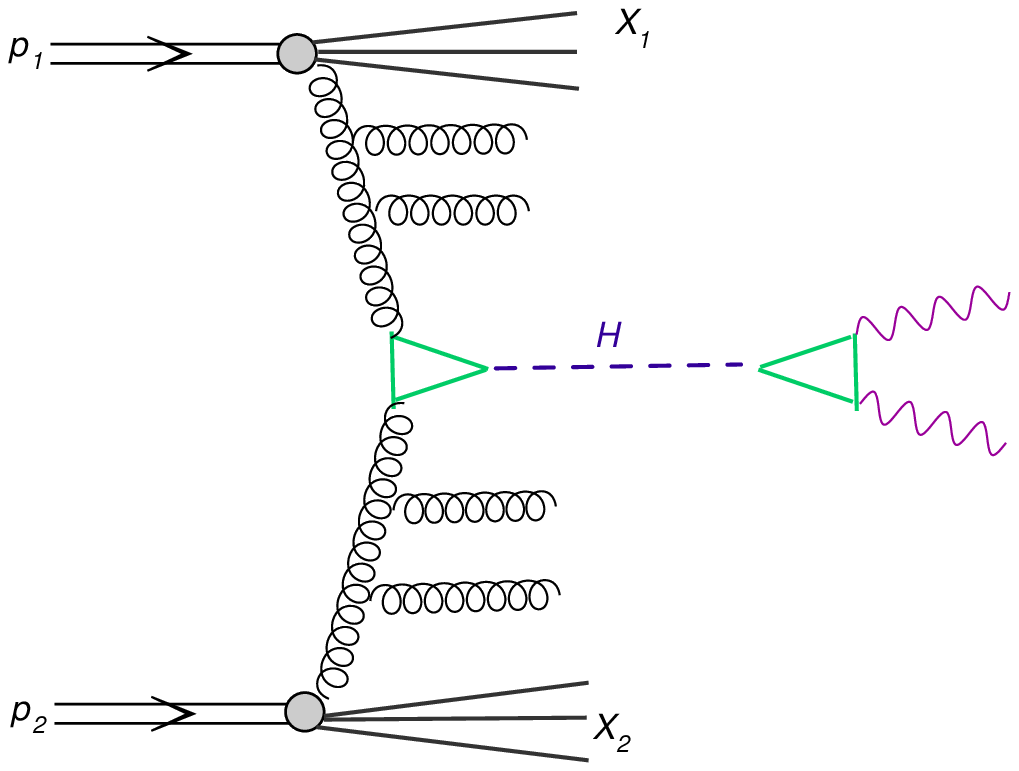}
\end{center}
{\caption 
\small The leading-order diagram for Higgs boson production
in the two-photon channel relevant for the $k_t$-factorization approach.
}
\label{fig:LO}
\end{figure}

The leading-order mechanism of Higgs boson production in 
the $k_t$-factorization is shown in Fig.\ref{fig:LO}.

In the $k_t$-factorization approach the cross section for 
the Higgs boson production can be written somewhat formally as:
\begin{eqnarray}
\sigma_{pp \to H} = \int \frac{dx_1}{x_1} \frac{dx_2}{x_2}
\frac{d^2 q_{1t}}{\pi} \frac{d^2 q_{2t}}{\pi} 
&&\delta \left( (q_1 + q_2)^2 - M_H^2 \right) 
\sigma_{gg \to H}(x_1,x_2,q_{1},q_{2}) \nonumber \\
&&\times \; {\cal F}_g(x_1,q_{1t}^2,\mu_F^2) {\cal F}_g(x_2,q_{2t}^2,\mu_F^2)
\; ,
\label{Higgs_kt_factorization}
\end{eqnarray}
where ${\cal F}_g$ are so-called unintegrated (or 
transverse-momentum-dependent) gluon distributions
and $\sigma_{g g \to H}$ is $g g \to H$ (off-shell) cross section.

\begin{figure*}
\begin{center}
\includegraphics[width=4.5cm]{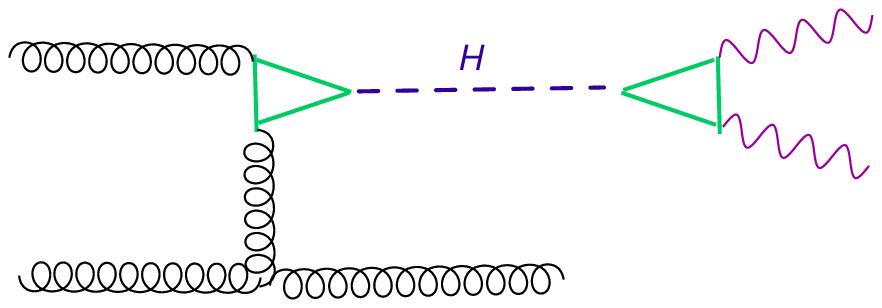}
\includegraphics[width=4.5cm]{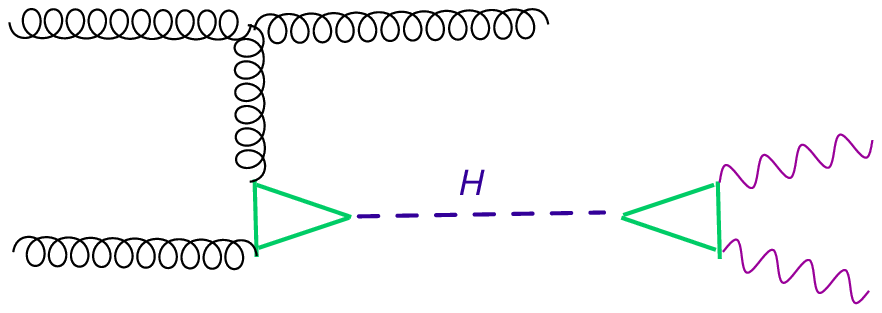}
\includegraphics[width=4.5cm]{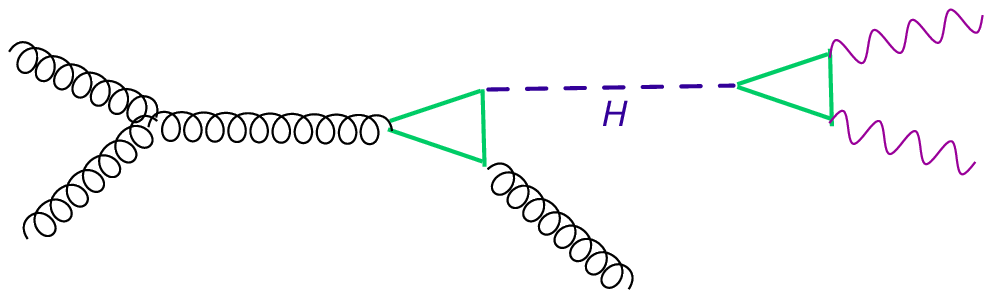}\\
\includegraphics[width=5cm]{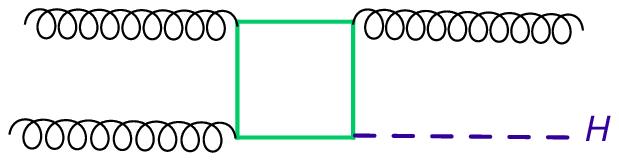}
\includegraphics[width=5cm]{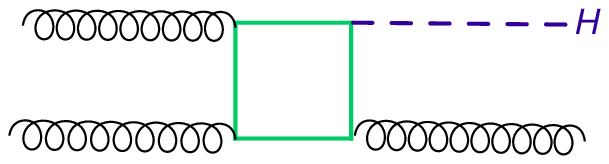}
\end{center}
{\caption 
\small QCD NLO subprocesses with triangles (upper line) and boxes
(lower line).
}
\label{fig:NLO}
\end{figure*}

In Fig.\ref{fig:NLO} we show next-to-leading order partonic
subprocesses. While the processes with triangles are effectively
included in a calculation related to the diagram shown in Fig.\ref{fig:LO},
the diagrams with boxes (much larger contributions) have to be
included extra.

In the collinear approximation the corresponding cross section 
differential in Higgs boson rapidity ($y_H$),
associated parton rapidity ($y_p$) and transverse momentum of each
of them can be written as:
\begin{eqnarray}
\frac{d \sigma}{dy_H dy_p d^2 p_t}(y_H,y_p,p_t) &=&
\frac{1}{16 \pi^2 {\hat s}^2 }  
\times \biggl\{ x_1 g_1(x_1,\mu^2)  x_2 g_2(x_2,\mu^2) 
 \overline{|{\cal M}_{gg \to Hg}|^2}   \nonumber \\   
&&+ \left[ \sum_{{f_1}=-3,3} x_1 q_{1,f_1}(x_1,\mu^2) \right] 
x_2 g_2(x_2,\mu^2)  \;
 \overline{|{\cal M}_{qg \to Hq}|^2}   \nonumber \\ 
&& + \; x_1 g_1(x_1,\mu^2)  
 \displaystyle \left[ \sum_{{f_2}=-3,3} x_2 q_{2,f_2}(x_2,\mu^2) \right]       
 \overline{|{\cal M}_{gq \to Hq}|^2}   \nonumber \\ 
&&+ \sum_{f=-3,3} x_1 q_{1,f}(x_1,\mu^2)  x_2 q_{2,-f}(x_2,\mu^2)
 \overline{|{\cal M}_{qq \to Hg}|^2}  \biggr\} .   
\label{2to2}
\end{eqnarray}
The indices $f$ in the formula above number both quarks ($f >$ 0)
and antiquarks ($f <$ 0). Only three light flavours are included in
actual calculations here.

In Ref.\cite{SLM2014} we have calculated the dominant $g g \to H g$
contribution also taking into account transverse momenta of initial gluons.
In the $k_t$-factorization the NLO differential cross section can 
be written as:
\begin{eqnarray}
\frac{d \sigma(p p \to H g X)}{d y_H d y_g d^2 p_{H,t} d^2 p_{g,t}}
&& = 
\frac{1}{16 \pi^2 {\hat s}^2} \int \frac{d^2 q_{1t}}{\pi} \frac{d^2 q_{2t}}{\pi} 
\overline{|{\cal M}_{g^{*} g^{*} \rightarrow H g}^{off-shell}|^2} 
\nonumber \\
&& \times \;\; 
\delta^2 \left( \vec{q}_{1t} + \vec{q}_{2t} - \vec{p}_{H,t} - \vec{p}_{g,t} \right)
{\cal F}(x_1,q_{1t}^2,\mu^2) {\cal F}(x_2,q_{2t}^2,\mu^2) \; .
\label{kt_fact_gg_Hg}
\end{eqnarray}
How the matrix elements were calculated was explained in our original
paper \cite{SLM2014}.

Also production of Higgs boson associated with two jets in the final state 
may play important role \cite{SLM2014}. 

\begin{figure*}
\begin{center}
\includegraphics[width=4.5cm]{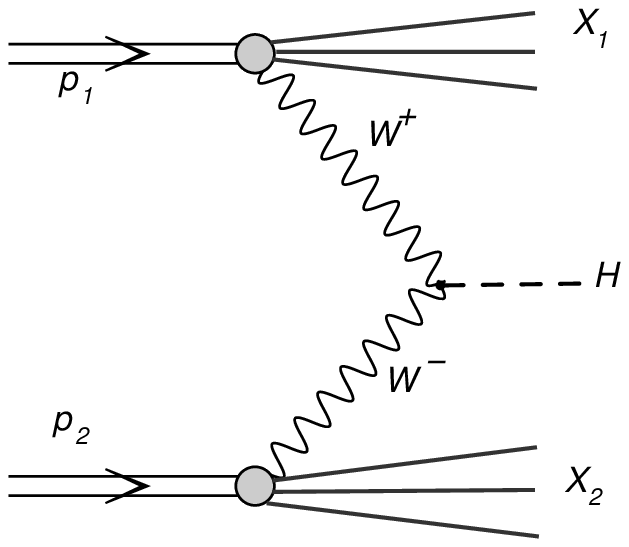}
\includegraphics[width=4.5cm]{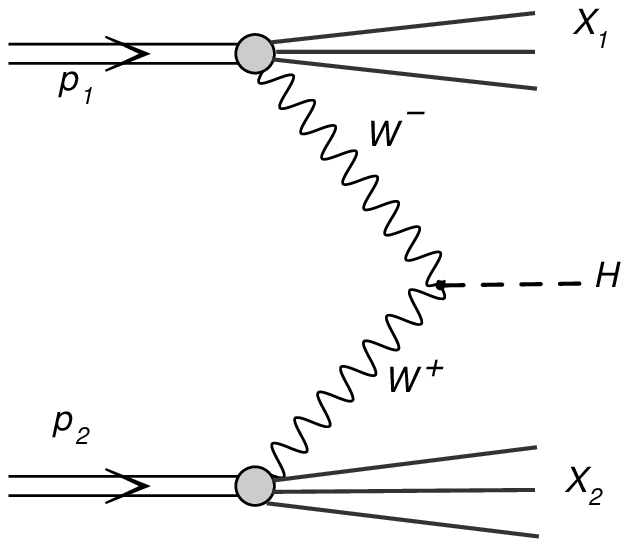}
\end{center}
{\caption 
\small Some examples of electoweak corrections with exchange of $W$
bosons
}
\label{fig:electroweak}
\end{figure*}

In addition to the QCD contributions discussed above one has to include
also electroweak ones. Some examples are shown in Fig.\ref{fig:electroweak}.

The corresponding proton-proton cross section can be written as
\begin{eqnarray}
d \sigma = {\cal F}_{12}^{VV}(x_1,x_2) \; \frac{1}{2 \hat s}
\; \overline{ | {\cal M}_{qq \to qqH} |^2 } \;
\frac{d^3 p_3}{(2 \pi)^3 2 E_3}
\frac{d^3 p_4}{(2 \pi)^3 2 E_4}
\frac{d^3 p_H}{(2 \pi)^3 2 E_H} \nonumber \\
\times \; (2 \pi)^4 \delta^{4}(p_1+p_2-p_3-p_4-p_H) \; d x_1 d x_2 \; .
\label{WW_fusion}
\end{eqnarray}
%

\section{Results}

\begin{figure}[!h]
\begin{center}
\includegraphics[width=5.5cm]{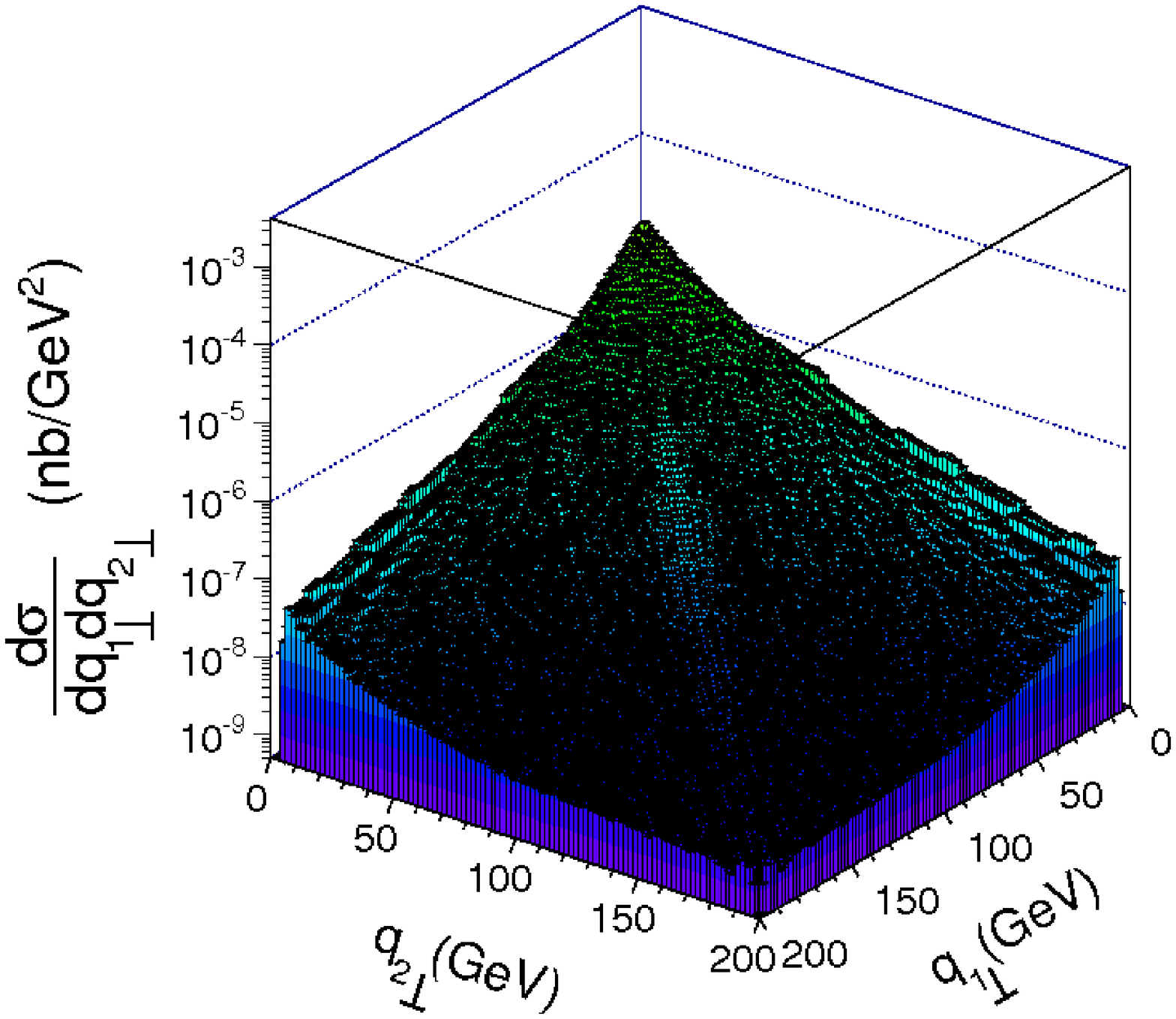}
\includegraphics[width=5.5cm]{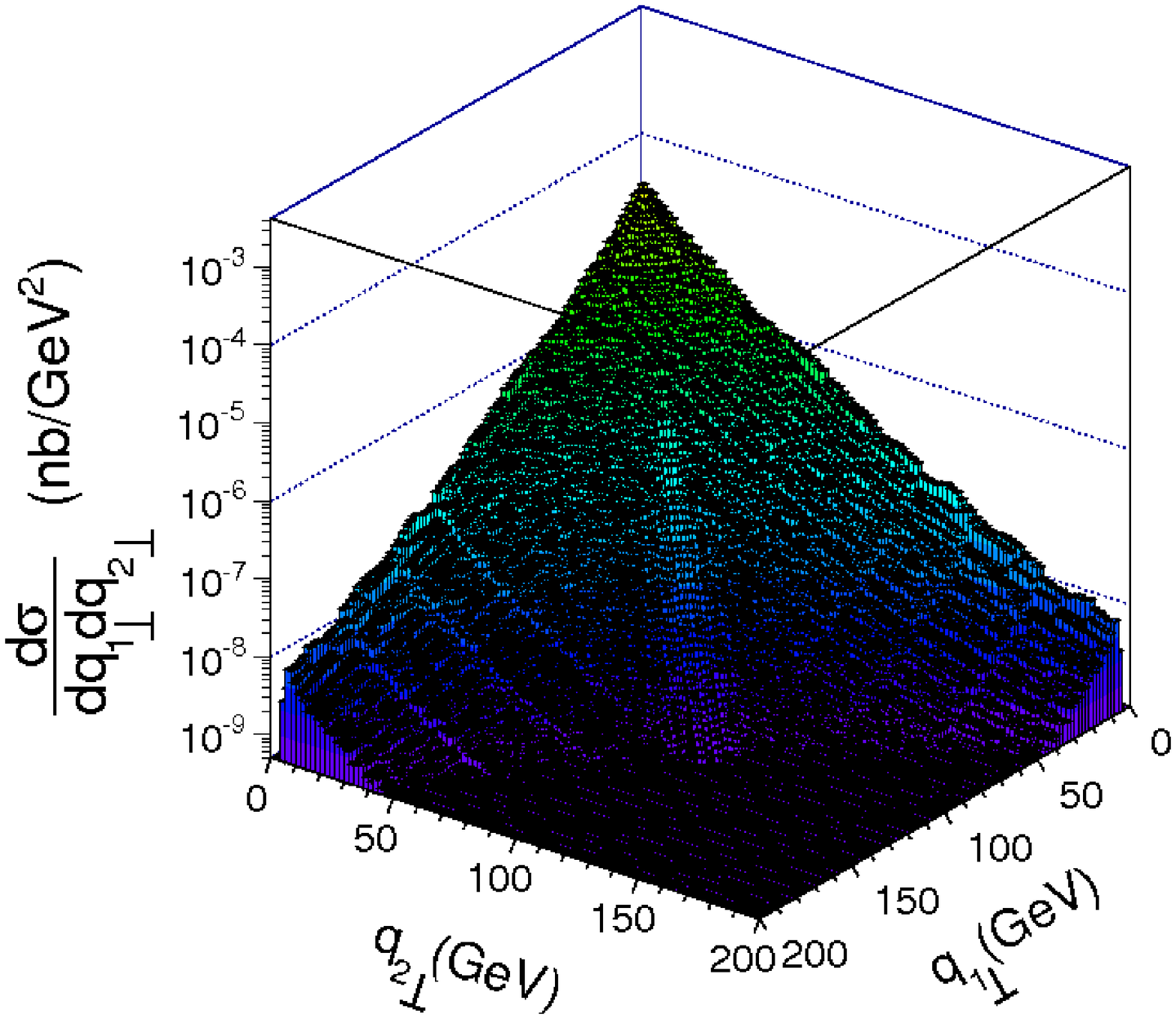}
\end{center}
   \caption{
\small Distribution in $q_{1t}$ and $q_{2t}$ for the $gg \to H$
subprocess and for the KMR and Jung CCFM (set$A0$) UGDFs.
}
 \label{fig:q1tq2t}
\end{figure}

The different UGDFs in the literature have quite different dependence 
on gluon transverse momenta. In Fig.~\ref{fig:q1tq2t} we show 
an example of two-dimensional maps in $q_{1t} \times q_{2t}$
(transverse momenta of the fusing gluons) for two UGDFs. 
Many more examples were presented in \cite{SLM2014}.
In Ref.\cite{SLM2014} we concluded that a use of saturation inspired
scale independent UGDFs is not sufficient for the Higgs boson
production. They lead also to distributions in $p_{1t}$ and $p_{2t}$
which are peaked in extremely small $p_{1t}$ and $p_{2t}$, at least
for the leading order $g g \to H$ subprocess.
Quite large gluon transverse momenta ($q_{1t},q_{2t} \sim m_H$) enter 
the production of the Higgs boson for the KMR and Jung CCFM (set$A0$) UGDFs. 
For the KMR UGDF a clear enhancement at small $q_{1t}$ or $q_{2t}$
can be observed. 
This is rather a region of nonperturbative nature, where the KMR UGDF 
is rather extrapolated than calculated.
We have checked, however, that the contribution of the region
when $q_{1t} <$ 2 GeV or $q_{2t} <$ 2 GeV constitutes only less than $5 \%$
of the integrated cross section. This is then a simple estimate
of uncertainty of the whole approach.

\begin{figure}[!h]
\begin{center}
\includegraphics[width=8cm]{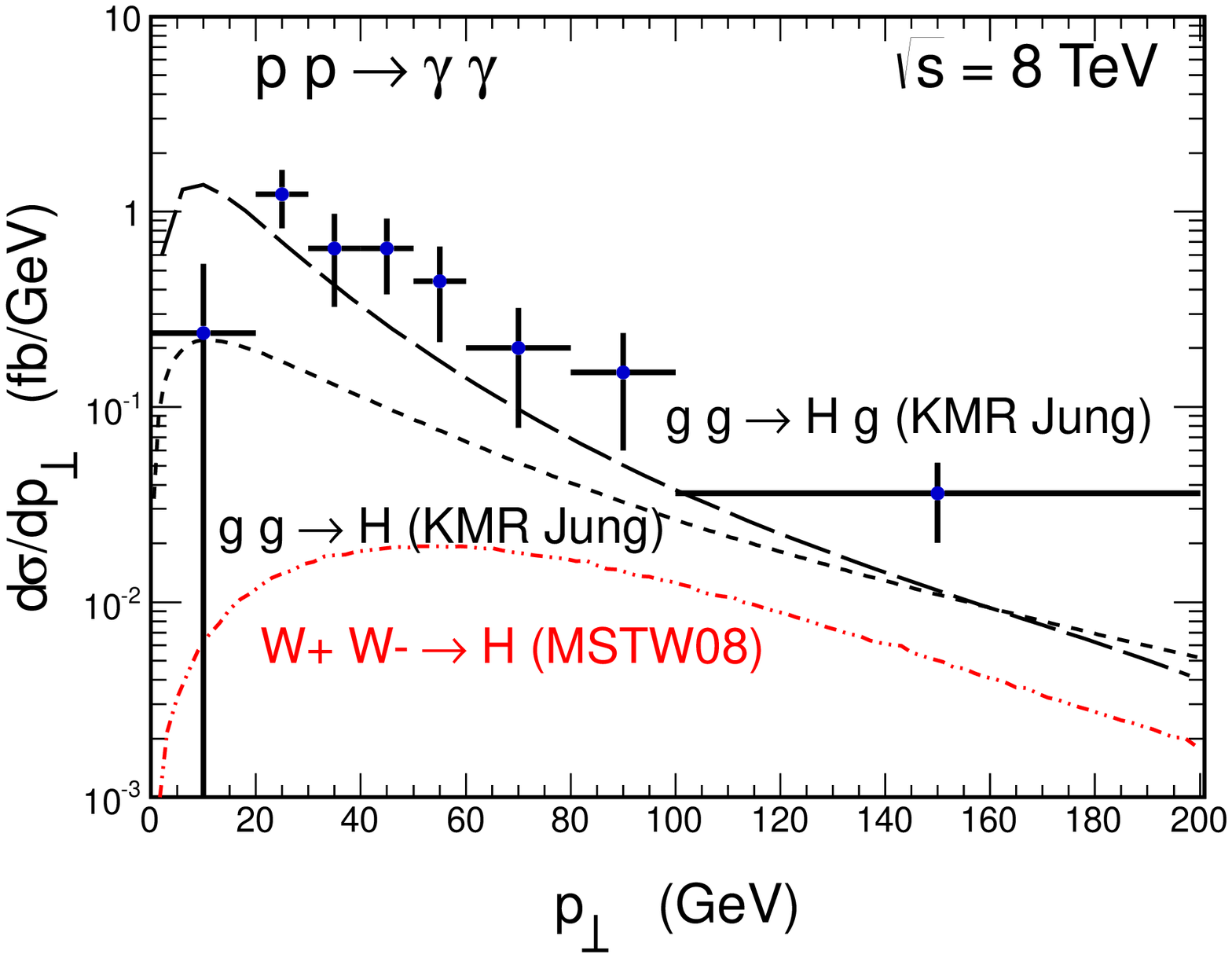}
\end{center}
   \caption{
\small Transverse momentum distribution of the Higgs boson
in the $\gamma \gamma$ channels for the different 
mechanisms: $g g \to H$ (solid line), $g g \to H g$ (dashed line) and 
$W W \to H$ (dash-dotted line).
}
 \label{fig:pt_gamgam_mechanisms}
\end{figure}

A distribution in Higgs boson transverse momentum is particularly
interesting.
In Fig.~\ref{fig:pt_gamgam_mechanisms} we compare contributions of
different mechanisms. The QCD contributions shown in this subsection 
were calculated with the KMR UGDF.
Surprisingly the contribution of the next-to-leading order 
mechanism $g g \to H g$ is even slightly bigger than that
for the $g g \to H$ fusion, especially for intermediate Higgs boson transverse
momenta. As discussed in our original paper \cite{SLM2014} there is
almost no double counting when adding the corresponding cross sections 
due to quite different topology of corresponding Feynman diagrams.
As shown in the present analysis the $g g \to H$ mechanism is not sufficient 
within the $k_t$-factorization approach.
The $2 \to 3$ contribution of the $g g \to g H g$ subprocess is also
not negligible but here one can expect that a big part is already 
contained in the $gg \to H$ calculation especially with the KMR UGDF. 
Therefore we do not add this contribution explicitly when
calculating $d\sigma/dp_{t,sum}$.
The contribution of the $WW$, $ZZ$ fusion is also fairly sizeable.
In principle, the Higgs bosons (or photons from the Higgs boson)
associated with the electroweak boson exchanges could 
be to some extend separated by requiring rapidity gaps i.e. production 
of Higgs boson isolated off other hadronic activity.


\section{Conclusions}

We have presented results of our analysis of production of Higgs boson 
in the two-photon channel within $k_t$-factorization. 
Matrix elements and UGDFs are the ingredients of the approach.

We have found that different UGDFs (not discussed here) give quite 
different results. However, many of the UGDF models were adjusted to low-x
phenomena and cannot be used for production of relatively heavy Higgs
boson, where rather large gluon transverse momenta are involved.

LO $k_t$-factorization underpredicts the two-photon
data in contrast to recent claims.
NLO corrections have to be taken into account
and the $g g \to H g$ subprocess is especially important.
Also a contribution of Higgs boson associated with quark/antiquark 
dijets is nonnegligible \cite{SLM2014}.
Electroweak corrections (here only $W^+ W^-$ fusion was discusssed) were
found to be large at large transverse momenta of the Higgs boson.
     
Only combined analysis including all ingredients can provide a
possibility to describe experimental data and to test UGDFs. 
We expect that future run2 data will allow for better tests of UGDFs.



\begin{thebibliography}{99}


\bibitem{Higgs_discovery}
G.~Aad {\it et al.} (the ATLAS Collaboration), Phys. Lett. {\bf B716}, 1 (2012);\\
S.~Chatrchyan {\it et al.} (the CMS Collaboration), Phys. Lett. {\bf B716}, 30 (2012).

\bibitem{Aad:2013wqa} 
  G.~Aad {\it et al.}  (the ATLAS Collaboration),
  Phys.\ Lett.\ {\bf B726}, 88 (2013); corrigendum: Phys.\ Lett.\ {\bf B734}, 406 (2014).

\bibitem{Chatrchyan:2013mxa} 
  S.~Chatrchyan {\it et al.}  (the CMS Collaboration),
  Phys.\ Rev.\ {\bf D89}, 092007 (2014).

\bibitem{Khachatryan:2014iha} 
  V.~Khachatryan {\it et al.}  (the CMS Collaboration),
  arXiv:1405.3455 [hep-ex].

\bibitem{Aad:2014aba} 
  G.~Aad {\it et al.}  (the ATLAS Collaboration),
  arXiv:1406.3827 [hep-ex].

\bibitem{NNLO}
R.V. Harlander and W.B. Kilgore,
Phys. Rev. Lett. {\bf 88} (2002) 201801;\\
C. Anastasiou and K. Melnikov,
Nucl. Phys. {\bf B646} (2002) 220;\\
V. Ravindran, J. Smith and W.L. van Neerven,
Nucl. Phys. {\bf B665} (2003) 325.



\bibitem{Jung2013}
P. Cipriano, S. Dooling, A. Grebenyuk, P. Gunnellini, F. Hautmann,
H. Jung and P. Katsas,
Phys. Rev. {\bf D88} (2013) 097501.

\bibitem{LMZ2014}
A.V. Lipatov, M.A. Malyshev and N.P. Zotov, Phys. Lett. {\bf B735}, 79 (2014); arXiv:1402.6481 [hep-ph].

\bibitem{KMR}
M.A. Kimber, A.D. Martin and M.G. Ryskin, Phys. Rev. {\bf D63} (2001) 114027;\\
G. Watt, A.D. Martin and M.G. Ryskin, Eur. Phys. J. {\bf C31} (2003) 73.

\bibitem{Jung}
H. Jung, G.P. Salam, Eur. Phys. J. {\bf C19} (2001) 351;\\
H. Jung, arXiv:0411287 [hep-ph].

\bibitem{ATLAS_Higgs}
ATLAS collaboration, ATLAS note, ATLAS-CONF-2013-072.

\bibitem{SLM2014}
A. Szczurek, M. Luszczak and R. Maciula,
Phys. Rev. {\bf D90} (2014) 094023.

































\end{thebibliography}
\end{document}